\documentclass[nouppercase]{ifmbe}

\usepackage{amsmath,amsfonts}
\usepackage{color, url}
\usepackage{fullpage}

\def\@{\hskip.8pt}
\def\?{\hskip.3pt}
\def\narc#1{\text{\@\tiny $(#1)$}}
\def \red#1 {\textcolor{red}{#1}}
\def \blue#1 {\textcolor{blue}{#1}}

\title{Bayesian multi--dipole localization and uncertainty quantification from simultaneous EEG and MEG recordings}

\affiliation{Dipartimento di Matematica, Universit\`a degli Studi di Genova}{FIRSTAFF}

\author{F. Rossi}{FIRSTAFF}
\author{G. Luria}{FIRSTAFF}
\author{S. Sommariva}{FIRSTAFF}
\author{A. Sorrentino}{FIRSTAFF}

\begin{document}

\maketitle

\begin{abstract}
We deal with estimation of multiple dipoles from combined MEG and EEG time--series.
We use a sequential Monte Carlo algorithm to characterize the posterior distribution of the number of dipoles and their locations. 
By considering three test cases, we show that using the combined data the method can localize sources that are not easily (or not at all) visible with either of the two individual data alone.
In addition, the posterior distribution from combined data exhibits a lower variance, i.e. lower uncertainty, than the posterior from single device.
\end{abstract}


\section{Introduction}

Electroencephalography (EEG) and Magnetoencephalography (MEG) provide non--invasive measurements of brain activity with high temporal resolution. 
The spatial resolution of these techniques, however, is partially limited by the ill--posedness of the inverse problem.
In particular, the two instruments are almost insensitive to currents orientated either radially (MEG) or tangentially (EEG). 
Therefore EEG and MEG recordings provide complementary information on the underlying neural activity \cite{malmivuo2012comparison,ahlfors2010sensitivity}.

For this reason, in the last two decades several methods have been proposed that combine EEG and MEG measurements using standard linear 
inversion \cite{babiloni2001linear,molins2008quantification}, sparse linear inversion \cite{ding2013simultaneous}, empirical Bayes with multiple sparse 
priors \cite{henson2009meg}, maximum entropy on the mean \cite{chowdhury2015meg}, beamformers \cite{ko2010beamformer}.
However, most of the existing studies that use simultaneous EEG and MEG data rely on a \textit{distributed} source model, rather than on the \textit{dipolar} one.
Recently we have been working on sequential Monte Carlo methods for approximating the posterior distribution of a set of current dipoles from M/EEG time--series \cite{sorrentino2009dynamical,soluar14,soso14,viso16}.
Here we consider a static multi--dipole model, in which 
the number of dipoles and their locations do not change with time. 
{This assumption is coherent with the neurophysiological interpretation of a current dipole as the activity of a neural population; in addition, it is commonly used in multi--dipole modeling of M/EEG data}.
%
%
We slightly modify the Sequential Monte Carlo algorithm developed in \cite{soso14}  in order to approximate the posterior distribution of the unknown parameters given simultaneous EEG and MEG recordings and show the advantages of the use of combined MEG and EEG data. 
Indeed, not only localization of purely radial/tangential sources becomes feasible, but also the uncertainty on the source location gets smaller.

The article is organized as follows. In Section II we review the main features of the algorithm originally proposed in \cite{soso14} for MEG data, and indicate how it can be modified to use simultaneous MEG and EEG data. In Section III we show three examples of application to synthetic time series. In Section IV we offer a brief discussion and our conclusions.

\section{Bayesian multi--dipole modeling from simultaneous EEG and MEG data}

\subsection{A Bayesian approach to multi--dipole modeling}

Let $y^\narc{E/M}_{1:T} = \left\{y_t^\narc{E/M}\right\}_{t=1,\dots,T}$ \vspace{1pt} be the recorded electric potential ($E$) or magnetic field ($M$), sampled at discrete time instants $t$.
In the classic multi--dipole model of M/EEG data, it is assumed that the recordings have been produced by a small set of point--like currents,
named \textit{current dipoles}, each of which is defined by its location $r$ and dipole moment $q$. The relationship between the measured data and the dipole parameters is given by

\begin{equation}
	y^\narc{E/M}_t = \sum_{n=1}^{N} G^\narc{E/M}(r^{\@n}) \cdot q^{\@n}_t +\, n^\narc{E/M}_t
	\label{eq:multi_dipole_model}
\end{equation}
where: $N$ is the number of dipoles; $G^\narc{E/M}(r)$, the \textit{leadfield}, is a known and non--linear function of the dipole location; 
$n^\narc{E/M}_t$ is zero--mean Gaussian noise, with covariance matrix $\Sigma^\narc{E/M}$. We remark that in our model the number of dipoles and their location do not change with time, while dipole moments do.

In a statistical Bayesian approach, one is interested in characterizing the posterior distribution for the unkonwn variables 
$x = \left\{N, r^1, q^1_{1:T}, \dots, r^N, q^N_{1:T}\@\right\}$, conditioned on the data:

\begin{equation}\label{eq:posterior}
	p(x\@|\@y^\narc{E/M}_{1:T}) = \frac{p(y^\narc{E/M}_{1:T} \@|\@x)\,p(x)}{p(\/y^\narc{E/M}_{1:T}\/)}\ .
\end{equation}

In \eqref{eq:posterior}, the likelihood function is the following Gaussian 

\begin{equation}
	p(y^\narc{E/M}_{1:T} \@|\@x) = \prod_t \mathcal{N} \left( y_t - \sum_{n=1}^{N} G^\narc{E/M}(r^n) \cdot q^n_t \; ; \; 0, \,  \Sigma^\narc{E/M} \right)\, ,
\end{equation}
while the prior distribution is 

\begin{equation}
	p(x) = p(N) \; \prod_{n=1}^N  \left( p(r^{\@n})\, \prod_{t=1}^T p(q_t^{\@n}) \right)
\end{equation}
where: $p(N)$ is chosen as a Poisson distribution with mean value $\lambda<1$, so as to favour low--cardinality models; $p(r^{\@n})$ is uniform in the brain volume;
$p(q^{\@n}_t)$ is Gaussian. 
In the present case, it is useful to split the posterior distribution as follows:
\begin{equation}
	p(x\@|\@y^\narc{E/M}_{1:T}) = p(q^{1:N}_{1:T}\@ |\@ r^{1:N},\,N,\, y_{1:T}^\narc{E/M})\ p(N,\,r^{1:N} \@|\@ y^\narc{E/M}_{1:T}) ~~~;
	\label{eq:posterior_split}
\end{equation}
indeed, under Gaussian assumptions for the prior $p(q^{1:N}_{1:T})$, the first bit at the right--hand  side is Gaussian and can be computed analytically, while
the second one can be approximated by means of Monte Carlo sampling.

\subsection{Adaptive sequential Monte Carlo sampling}

\textbf{Sequential Monte Carlo algorithm.} We use a Monte Carlo technique to approximate the posterior distribution for the number and location of dipoles, 
defined in equation (\ref{eq:posterior_split}). Specifically, we adopt a recently developed class of algorithms known as Sequential Monte Carlo samplers \cite{dedoja06} 
that behave very efficiently in the case of complex posteriors. The two main ideas underlying this class of algorithms are:
\begin{itemize}
  \item instead of trying to sample the posterior directly, they reach the posterior distribution smoothly,
  through a tempering sequence of auxiliary distributions. Here we adopt the sequence defined as
\begin{equation}
p_k(\@N,r^{\@1:N}\@|\@y_{1:T}^\narc{E/M}\@) \propto p(\@y_{1:T}^\narc{E/M} \@|\@ N,\,r^{\@1:N})^{\alpha(k)}\@ p(\@N,\,r^{\@1:N}\@)\, ,
\end{equation}
  with $\alpha(1)=0\@$, $\alpha(K)=1$ and $\alpha(1) < \alpha(2) < \ldots < \alpha(K)$, being $K$ the number of iterations; 
  \item they  combine Importance Sampling (IS) and Markov Chain Monte Carlo (MCMC) \cite{roca04}: first, they draw a given number $I$ of uniformly weighted
   samples from the prior distribution; then, they let each sample evolve, at every iteration, according to a Markov Chain Monte 
  Carlo transition kernel (that is $p_{k+1}$--invariant), and re--weight the samples appropriately. Such 
  combination of IS and MCMC crucially combines the benefits of IS (independent samples, global exploration
  of the state--space) with those of MCMC (correlated samples, good local exploration of the state--space).
\end{itemize}
At each iteration, the algorithm gets a sample set that is distributed according to $p_k(N,r^{1:N}\@|\@y_{1:T}^\narc{E/M})$; sample points can therefore account 
gradually for the information content of the data.
Importantly, the sequence of values of the exponent (and therefore the total number of iterations) needs not to be established a priori; instead, it can be adaptively determined at run--time, 
based on how well the sample set $\left\{(N_i,r^{\@1:N}_i\@)\right\}_{i=1,...,I}\@$ approximates $p_k(N,r^{\@1:N}\@|\@y_{1:T}^\narc{E/M})\@$.\\


\noindent
\textbf{Estimates.} The approximated posterior distribution contains information on multiple alternative models; therefore in order
to produce a sensible map, we restrict our attention to the most probable model, by first estimating
the posterior probabilities for the number of sources; i.e. we compute the $\arg\max \mathbb{P}(N\@|\@y_{1:T}^\narc{E/M})$, with
\begin{equation}
\mathbb{P}(N\@|\@y_{1:T}^\narc{E/M})\, = \,\sum_{i=1}^I w_i\, \delta(N,N_i)
\label{eq:mod_sel}
\end{equation}
being $w^i$ and $N^i$ the weight and the number of dipoles of the
i–-th sample respectively. Subsequently, denoting by $\hat{N}\@$ the most probable number of dipoles, for each voxel $r$, we compute 
\begin{equation}
\mathbb{P}(r\@|\@y_{1:T}^\narc{E/M},\@\hat{N})\, =\, \sum_{i=1}^I w_i\@ \delta(\hat{N},N_i)\, \sum_{k=1}^{N_i} \delta(r, r^k_i)\ ;
\label{eq:estimate}
\end{equation}
this quantity represents the posterior probability of there being a dipole in that voxel; on the other hand, the integral of $\mathbb{P}(r\@|\@y_{1:T}^\narc{E/M},\@\hat{N})$ over 
a volume can be interpreted as the expected number of sources in the integration volume.

\subsection{Combining MEG and EEG data}

Here we aim at characterizing the posterior distribution for the unknowns given simultaneous recordings of the electric potential and magnetic field. 
In fact, there will be no formal difference with the single--device scenario.
If we concatenate the data and define $y_t^\narc{E+M} = (\/y_t^\narc{E},\, y_t^\narc{M}\/)$, and do the same for the leadfield matrices 
$G^\narc{E+M} = [\@G^\narc{E} \@ G^\narc{M}]\@$, and the noise vectors $n^\narc{E+M}_t = (\/n^\narc{E}_t,\/ n^\narc{M}_t\/)$, then we can rewrite 
equation (\ref{eq:multi_dipole_model}) as:  
\begin{equation}
	y^\narc{E+M}_t = \sum_{n=1}^{N} G^\narc{E+M}(r^{\@n}) \cdot q^{\@n}_t\ +\ n^\narc{E+M}_t \ ,
	\label{eq:multi_dipole_model_EM}
\end{equation}
where, of course,  the underlying hypothesis is that the dipoles producing the measured EEG and MEG data are the same. 
Then one can go on with exactly the same discussion as before. 
The key point is to set the noise covariance matrix properly:  in the present case, we will be using

\begin{equation}
\Gamma^\narc{\mathbf{E+M}} = \left( \begin{array}{cc} \Gamma^\narc{E} & 0\\
                                   0 &  \Gamma^\narc{M} \end{array} \right)\ .
\end{equation}

\section{Results}

\begin{figure}[ht]
      \centering
          \includegraphics[width=0.95\columnwidth]{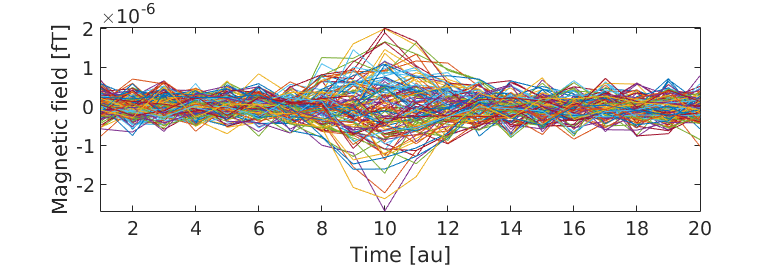}
      \caption{MEG dataset.}
      \label{fig:data}
\end{figure}

We consider three exemplar test cases: two single--dipole cases and one three--dipole case.
The leadfield matrices were generated with the BESA Research software. 
The EEG leadfield contains 32 sensors, the MEG leadfield contains 122 sensors. The source space is discretized with 18,023 voxels (possible dipole locations). 
{The electrical conductivity inside the head was modeled as piecewise constant and homogeneous; a realistic head geometry was used; the leadfield calculation was performed using a boundary element method.}
Synthetic MEG and EEG data were generated by choosing dipole location(s) and orientation(s), and then assigning dipoles
a bell--shaped source time course. White Gaussian noise was added to the synthetic data to produce a Signal--to--Noise Ratio
comparable to that of experimental data. The MEG synthetic data generated for the three--dipole case are shown in Figure \ref{fig:data}.

In the two single--dipole cases, dipole orientations were selected to make the dipole tangential, in one case, and radial, in the other case, so as to make localization 
from MEG or EEG only, respectively, particularly challenging.

\begin{figure}[ht]
      \centering
          \includegraphics[width=0.95\columnwidth]{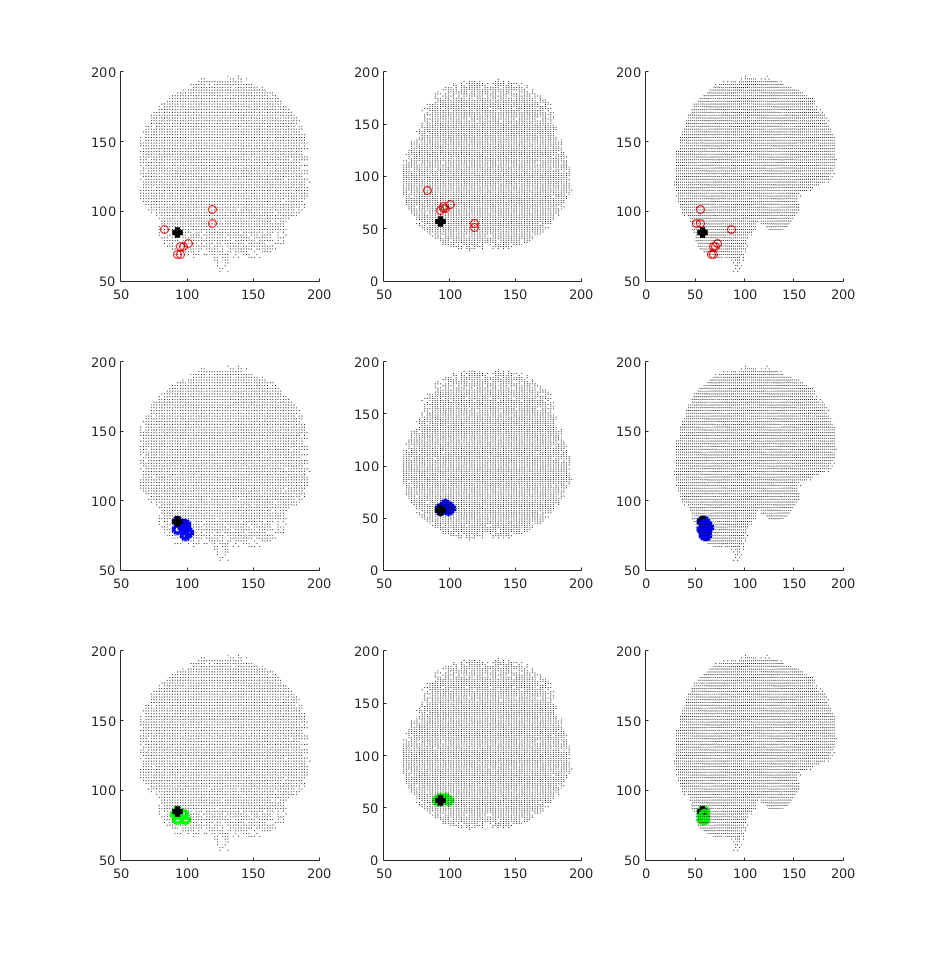}
      \caption{One tangential dipole. Red: EEG. Blue: MEG. Green: EEG+MEG.}
      \label{fig:1dip1}
\end{figure}

\begin{figure}[ht]
      \centering
          \includegraphics[width=0.95\columnwidth]{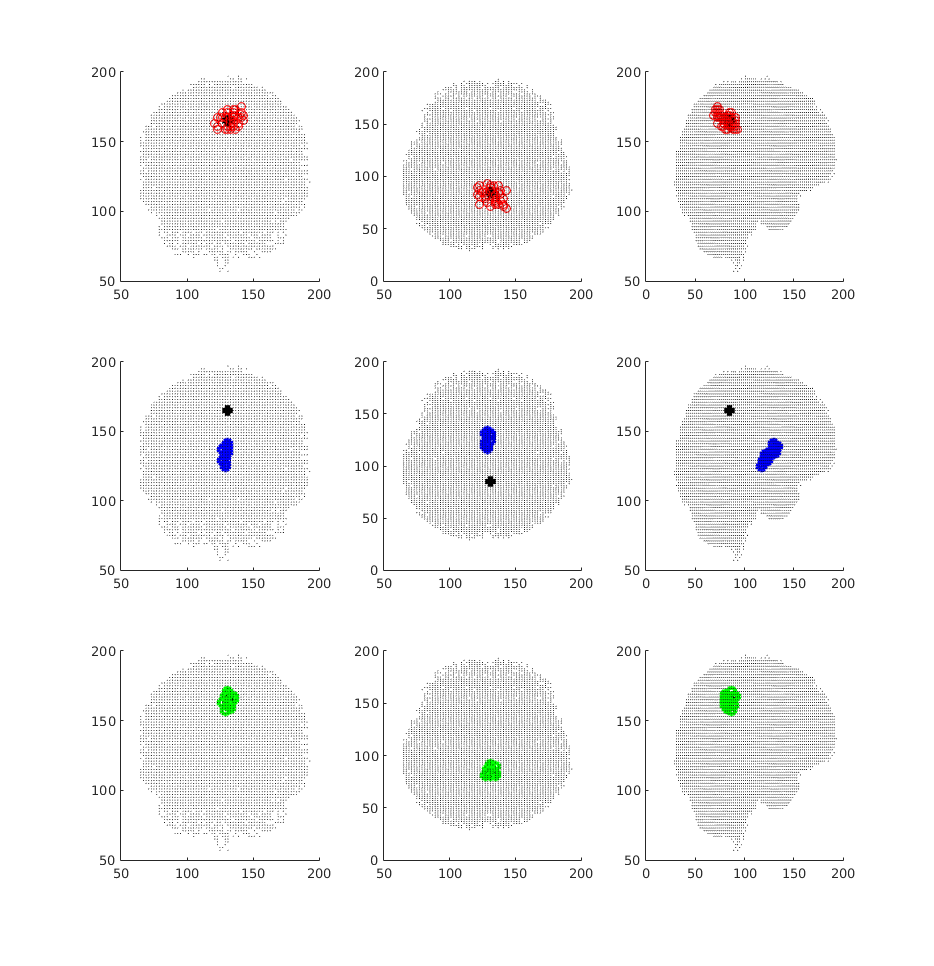}
      \caption{One radial dipole. Red: EEG. Blue: MEG. Green: EEG+MEG.}
      \label{fig:1dip2}
\end{figure}

\begin{figure}[ht]
      \centering
          \includegraphics[width=0.95\columnwidth]{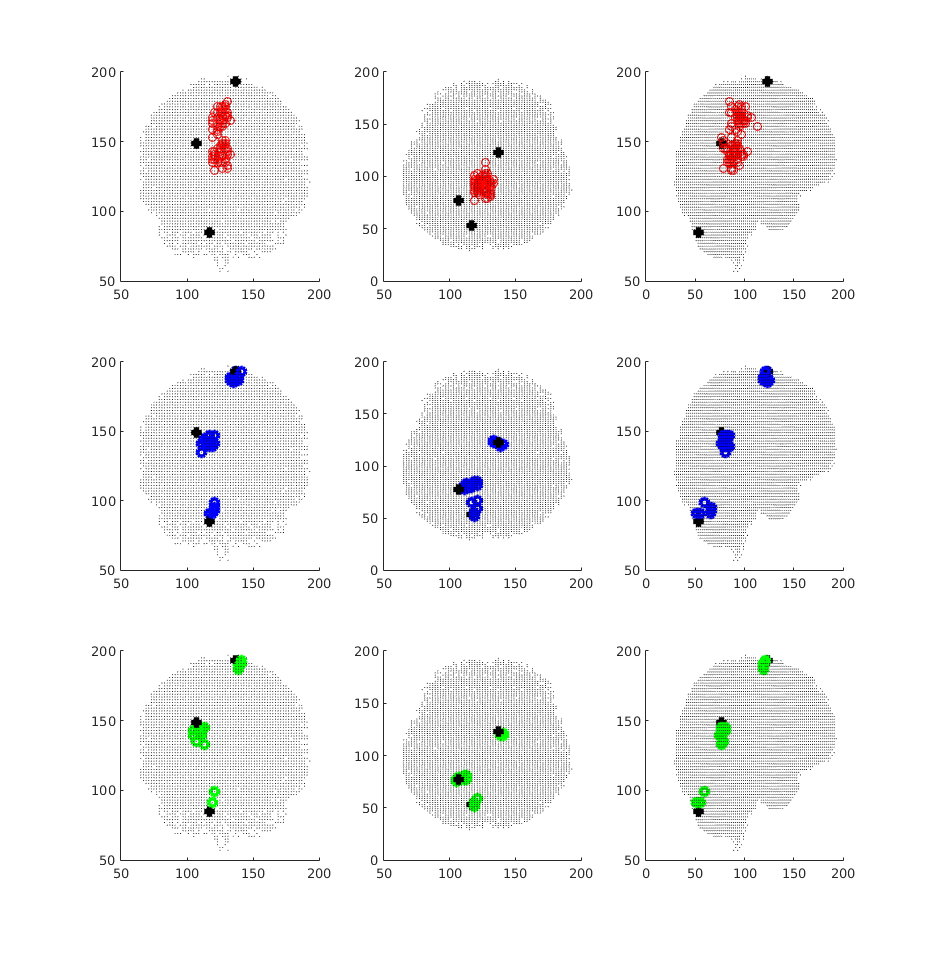}
      \caption{Three dipole case. Red: EEG. Blue: MEG. Green: EEG+MEG.}
      \label{fig:3dip}
\end{figure}

For each test dataset, we applied the SMC algorithm to the EEG data only, to the MEG data only and to the combined EEG+MEG data.
In Figures \ref{fig:1dip1}--\ref{fig:3dip} we show the localization results: we highlight the points having empirical posterior probability, 
as computed by (\ref{eq:estimate}) above the threshold $P_{th} = 1/1000$. Red, blue and green refer to the results obtained from the EEG data only, 
from MEG data only, and from combined EEG+MEG, respectively.

As expected, even single dipole localization is challenging, with M/EEG only, when the dipolar sources are tangential/radial. 
For the tangential source (Fig. \ref{fig:1dip1}), the posterior from EEG data is smeared and indicates several possible distinct locations.
For the radial source (Fig. \ref{fig:1dip2}), the posterior from MEG data concentrates in a deep and central area of the brain, confirming that the signal does not allow proper localization of the source. However, in both cases the posterior from the combined data is able to localize the dipolar source. 
{In order to quantify the improvement, here we compute the average distance from the true dipole location $r^*$, i.e.:} $ \Delta = E_{\mathbb{P}(r|y_{1:T},\hat{N})} \|r-r^{*}\|$.
{We notice that this quantity contains information on both accuracy of the reconstruction and variance of the distribution. For the tangential dipole, we get $\Delta^{(E)} = 28$ mm, $\Delta^{(M)} = 8$ mm and $\Delta^{(E+M)} = 0.4$ mm; for the radial dipole, we get $\Delta^{(E)} = 7$ mm, $\Delta^{(M)} = 50$ mm and $\Delta^{(E+M)} = 0.4$ mm, again. Remarkably, in both cases the results from the combined measurement are better than those from the best single--measurement case.}

For the three--dipole case (Fig. \ref{fig:3dip}), the EEG posterior indicates two sources, none of which corresponds exactly to one of the three true dipoles.
The MEG localization is quite accurate, however, also in this case the combined localization is better, featuring a smaller uncertainty. 
{Quantitative evaluation of the performance is more delicate in this case, and is left for a more thorough study.}

\section{Discussion and Conclusions}

The preliminary results contained in this article confirm that the combination of EEG and MEG data can provide better localization
with respect to the use of just one device. Interestingly, the use of a Bayesian approach in this context allows to characterize the performance improvement not only in terms of missing sources or localization error, but also in terms of uncertainty of the estimated location, coded in the variance of the posterior probability distribution. 

From a statistical perspective, the results of this paper cannot be considered as a fair comparison between EEG, MEG and EEG+MEG, because the number of sensors considered in the three cases is not the same. At the same time, the results can be considered of practical relevance, inasmuch as the comparison has been performed using typical sensor configurations, as available in many research centers and clinics.

Future work will be devoted to investigate this whole problem more systematically, by extending the number of test cases and providing quantitative measures of improvements.

\section*{Conflict of Interest}
The authors declare that they have no conflict of interest.

\section*{Acknowledgements}

BESA GmbH, and particularly Andre Waelkens, is kindly acknowledged for providing the leadfield matrix.

\bibliography{biblio}

\end{document}